\documentclass[letterpaper]{ptephy}
\preprintnumber{XXXX-XXXX}

\usepackage{boites}
\usepackage{amsmath}	

\usepackage{amsmath}
\usepackage{amssymb}
\usepackage{bm}
\usepackage[subrefformat=parens]{subcaption}

\makeatletter
\DeclareRobustCommand{\cev}[1]{%
  \mathpalette\do@cev{#1}%
}
\newcommand{\do@cev}[2]{%
  \fix@cev{#1}{+}%
  \reflectbox{$\m@th#1\vec{\reflectbox{$\fix@cev{#1}{-}\m@th#1#2\fix@cev{#1}{+}$}}$}%
  \fix@cev{#1}{-}%
}
\newcommand{\fix@cev}[2]{%
  \ifx#1\displaystyle
    \mkern#23mu
  \else
    \ifx#1\textstyle
      \mkern#23mu
    \else
      \ifx#1\scriptstyle
        \mkern#22mu
      \else
        \mkern#22mu
      \fi
    \fi
  \fi
}

\usepackage{version}	

\usepackage{color}
\newcommand{\TF}[1]{#1}
\newcommand{\KS}[1]{#1}
\newcommand{\KK}[1]{#1}
\newcommand{\TT}[1]{#1}
\newcommand{\SK}[1]{#1}
\usepackage{ulem}
\begin{document}
\title{
Large-amplitude quadrupole shape mixing probed by the $(p,p^\prime)$ reaction
: a model analysis
}

\author{\name{\fname{Koichi} \surname{Sato}}{1,2,*} 
\name{\fname{Takenori} \surname{Furumoto}}{3}, 
\name{\fname{Yuma} \surname{Kikuchi}}{4},
\name{\fname{Kazuyuki} \surname{Ogata}}{1, 2, 5},
\name{\fname{Yukinori} \surname{Sakuragi}}{1,2}
}

\address{%
\affil{1}{Department of Physics, Osaka City University, Osaka 558-8585, Japan}
\affil{2}{Nambu Yoichiro Institute of Theoretical and Experimental Physics (NITEP), Osaka City University, Osaka 558-8585, Japan}
\affil{3}{Graduate School of Education, Yokohama National University, Yokohama 240-8501, Japan}
\affil{4}{National Institute of Technology, Tokuyama College, Shunan 745-8585, Japan}
\affil{5}{Research Center for Nuclear Physics (RCNP), Osaka University, Ibaraki 567-0047, Japan}
\email{satok@sci.osaka-cu.ac.jp}
}

%

\begin{abstract}
To discuss a possible observation of large-amplitude nuclear shape \KK{mixing}  \KK{by}
 nuclear reaction,
we employ a simple collective model and evaluate transition
 densities, with which the differential cross sections are obtained
 through the microscopic coupled-channel calculation.
\KK{Assuming the spherical-to-prolate shape transition, we focus on
large-amplitude shape mixing associated with the softness of the
 collective potential in the $\beta$ direction.}
We introduce a simple model based on the five-dimensional quadrupole
collective Hamiltonian, which simulates
a chain of isotopes that exhibit spherical-to-prolate shape phase transition.
Taking $^{154}$Sm as an example and controlling the model parameters,
we study how the large-amplitude shape mixing affects the elastic and
 inelastic proton scatterings.
%
The calculated results suggest that \TF{the inelastic cross section of the $2_2^+$ state tells us an important role of the quadrupole shape mixing.}
\end{abstract}

\subjectindex{xxxx, xxx}

\maketitle
In low-lying states in atomic nuclei quadrupole deformation
plays an important role. \KK{The collectivity associated with the nuclear quadrupole deformation}
is experimentally studied through excitation energies, $B(E2)$, 
spectroscopic quadrupole moments and so on. 
Theoretically, 
one of the standard tools to describe the quadrupole deformation
and rotation dynamics is
the five-dimensional (5D) quadrupole collective Hamiltonian
\cite{Bohr1975, Belyaev1965, Kumar1967, Prochniak2009, Matsuyanagi2016}, 
and it is widely used in nuclear structure studies.
The dynamical variables in this 5D collective Hamiltonian approach are the 
magnitude and triaxiality of quadrupole deformation $(\beta,\gamma)$,
and the three Euler angles. \KK{The collective Hamiltonian is
characterized by the collective potential and inertial masses, which are
introduced either phenomenologically or microscopically.}
By solving the collective Schr\"odinger equation, 
one can describe \KK{large-amplitude} quadrupole shape mixing dynamics as well as
three-dimensional nuclear rotation.

\SK{There exist some preceding studies on nuclear shape mixing dynamics
by means of electron and nucleon scatterings.}
\KK{The electron scattering
is a powerful tool to determine the charge distribution and transition
charge densities. 
In Ref. \cite{Phan1988,Boeglin1988,Sandor1991},
the collective Hamiltonian microscopically derived was used to evaluate the transition
densities including dynamical deformation. 
It is also worth mentioning here that, recently, Yao et al. proposed a method
of calculating form factors and transition densities within
the beyond-mean-field framework of particle-number and angular-momentum
projected generator coordinate method~\cite{Yao2015}.
In the analysis of the nucleon scattering experiments in the 1980s
\cite{Mirzaa1985,Delaroche1987,Clegg1989,Hicks1989}, 
several nuclear structure models were adopted in conjunction with the coupled-channel
calculation
to investigate the $\gamma$-softness in the low-lying excited states in
the Pt-Os region, and it was found that some isotopes are best described
by $\gamma$-soft models.}

In this Letter, \KK{we discuss such a possible observation of
large-amplitude shape mixing by nuclear reaction.}
We investigate
the effect of quadrupole shape mixing on the \TF{proton elastic and inelastic scatterings}
with use of a simple collective model based on the 5D quadrupole
collective Hamiltonian.
Here, we focus on shape mixing dynamics in the 
\KK{spherical-to-prolate transition},
assuming transitional nuclei in the $A \sim 150$ mass region, and apply our 
model to a samarium isotope.
The purpose of this study is, \KK{however,} 
to  grasp the gross feature of the effects of
the quadrupole deformation and large-amplitude 
shape mixing dynamics on the \TF{observable} differential cross sections
for the $(p, p^\prime)$ reaction from a general point of view.



In Ref. \cite{Sato2010},  the triaxial deformation dynamics was studied
by introducing the collective potential and inertial masses
in the collective Hamiltonian phenomenologically.
We follow this approach and extend the model proposed in Ref. \cite{Sato2010} to study
the case where the collective potential is soft against the $\beta$ deformation.
The 5D collective Hamiltonian is written as
\begin{align}
\mathcal{H}_{\rm coll}&=T_{\rm vib} +T_{\rm rot} +V_{\rm coll}(\beta,\gamma),
 \label{eq:classical H}\\
          T_{\rm vib} &=\frac{1}{2}D_{\beta\beta}\dot \beta^2
+D_{\beta\gamma}\dot \beta\dot \gamma+\frac{1}{2}D_{\gamma\gamma}\dot
 \gamma^2, \\
T_{\rm rot} &=\frac{1}{2}\sum_{k=1}^3\mathcal{J}_k(\beta,\gamma)\omega_k^2,
\end{align}
where $T_{\rm vib}$ and $T_{\rm rot}$ are the vibrational and rotational
kinetic energies, \SK{respectively,} and $V_{\rm coll}$ is the collective potential.
The 5D collective Hamiltonian is characterized by seven quantities:
the collective potential $V_{\rm coll}(\beta, \gamma)$, 
three vibrational masses $D_{\beta\beta}(\beta,\gamma),
D_{\beta\gamma}(\beta, \gamma),
D_{\gamma\gamma}(\beta, \gamma)$,
and three moments of inertia $\mathcal{J}_k(\beta, \gamma)~(k=1,2,3)$.
The collective potential $V_{\rm coll}(\beta, \gamma)$ must be a scalar under
rotation, so it can be written as a function of $\beta^2$ and
$\beta^3\cos 3\gamma$ \cite{Bohr1975}.
We 
consider three typical 
situations: a spherical vibrator, a prolate rotor, 
and a transitional nucleus between the former two limits.
To simulate those typical situations, 
we adopt the collective potential in the following form:
\begin{align}
 V_{\rm coll}(\beta,\gamma)
=&\frac{1}{2}C_2 \beta^2+\frac{1}{2}C_4\left(\beta^2-\beta^2_0\right)^2 
   +C_6\beta^6
+ v_1\beta^3\cos 3\gamma, \label{eq:collective potential}
\end{align}
which is a modification of the potential in Ref. \cite{Sato2010}.
The inertial mass parameters are the same as used in Ref. \cite{Sato2010}:
\begin{align}
 D_{\beta\beta}(\beta,\gamma)&=D\left(1-\epsilon^\prime \beta\cos
 3\gamma\right),\label{eq:Dbb}\\
 D_{\gamma\gamma}(\beta,\gamma)&=D\beta^2\left(1+\epsilon^\prime \beta\cos 3\gamma\right),\label{eq:Dgg}\\
 D_{\beta\gamma}(\beta,\gamma)&=D\epsilon^\prime \beta\sin 3\gamma,\label{eq:Dbg}\\
 D_{k}(\beta,\gamma)&=D\left(1+\epsilon^\prime \beta\cos\gamma_k \right)
\,\, (k=1,2,3), \label{eq:Dk}
\end{align}
where $\gamma_k=\gamma -(2\pi k)/3$, and the three moments of inertia
are given
by $\mathcal{J}_k(\beta,\gamma)=4\beta^2
D_k(\beta,\gamma)\sin^2\gamma_k$.
By controlling the parameters in Eqs. (\ref{eq:collective
potential})--(\ref{eq:Dk}),
we simulate the three typical situations mentioned above.
After quantizing the collective Hamiltonian Eq. (\ref{eq:classical H}),
we solve the collective Schr\"odinger equation and obtain the collective
wave functions
\begin{align}
\Psi_{ \alpha I M }(\beta,\gamma,\Omega) = \sum_{K:{\rm even}}
 \Phi_{\alpha I K }(\beta,\gamma) \langle \Omega|IMK\rangle . 
\end{align}

\begin{figure}[tb]
 \begin{minipage}[b]{0.32\linewidth}
  \centering
  \includegraphics[keepaspectratio, scale=0.2]{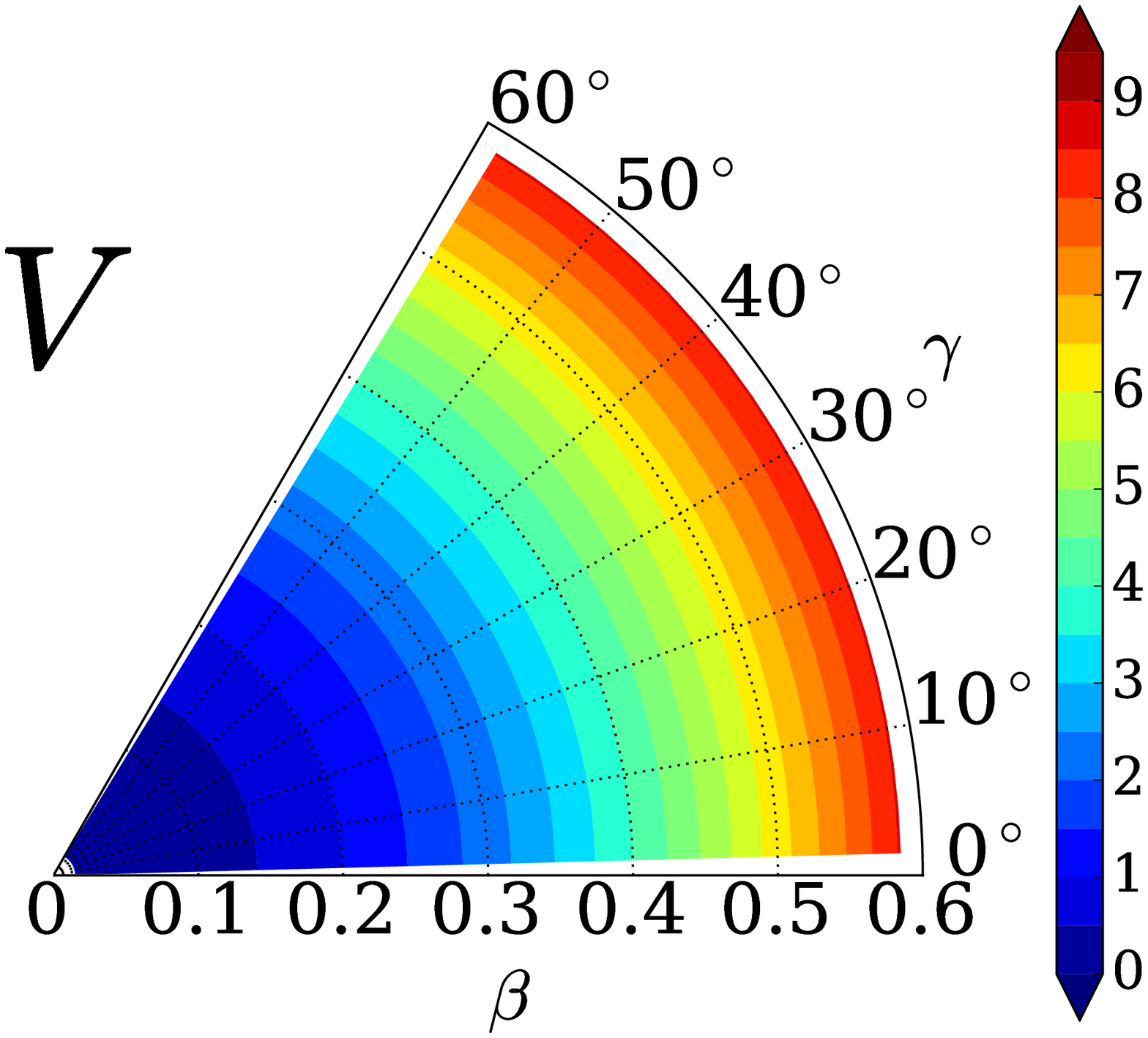}
  \vspace{0.5em}
  \subcaption{spherical vibrator}\label{V:HO}
 \end{minipage}
 \begin{minipage}[b]{0.32\linewidth}
  \centering
  \includegraphics[keepaspectratio,  scale=0.2]{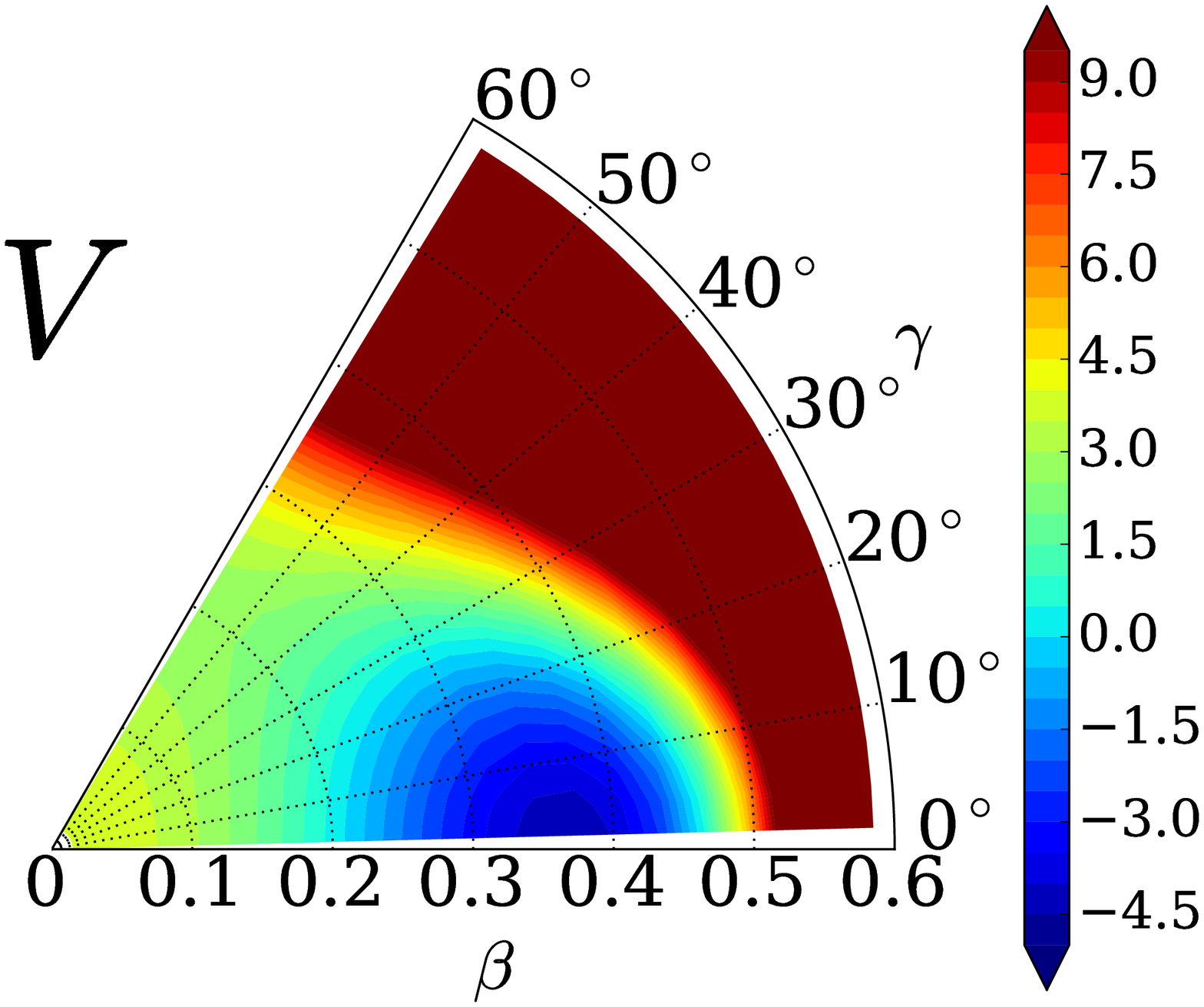}
  \vspace{0.5em}
  \subcaption{prolately-deformed
}\label{V:prolate}
 \end{minipage}
 \begin{minipage}[b]{0.32\linewidth}
  \centering
  \includegraphics[keepaspectratio,scale=0.2]{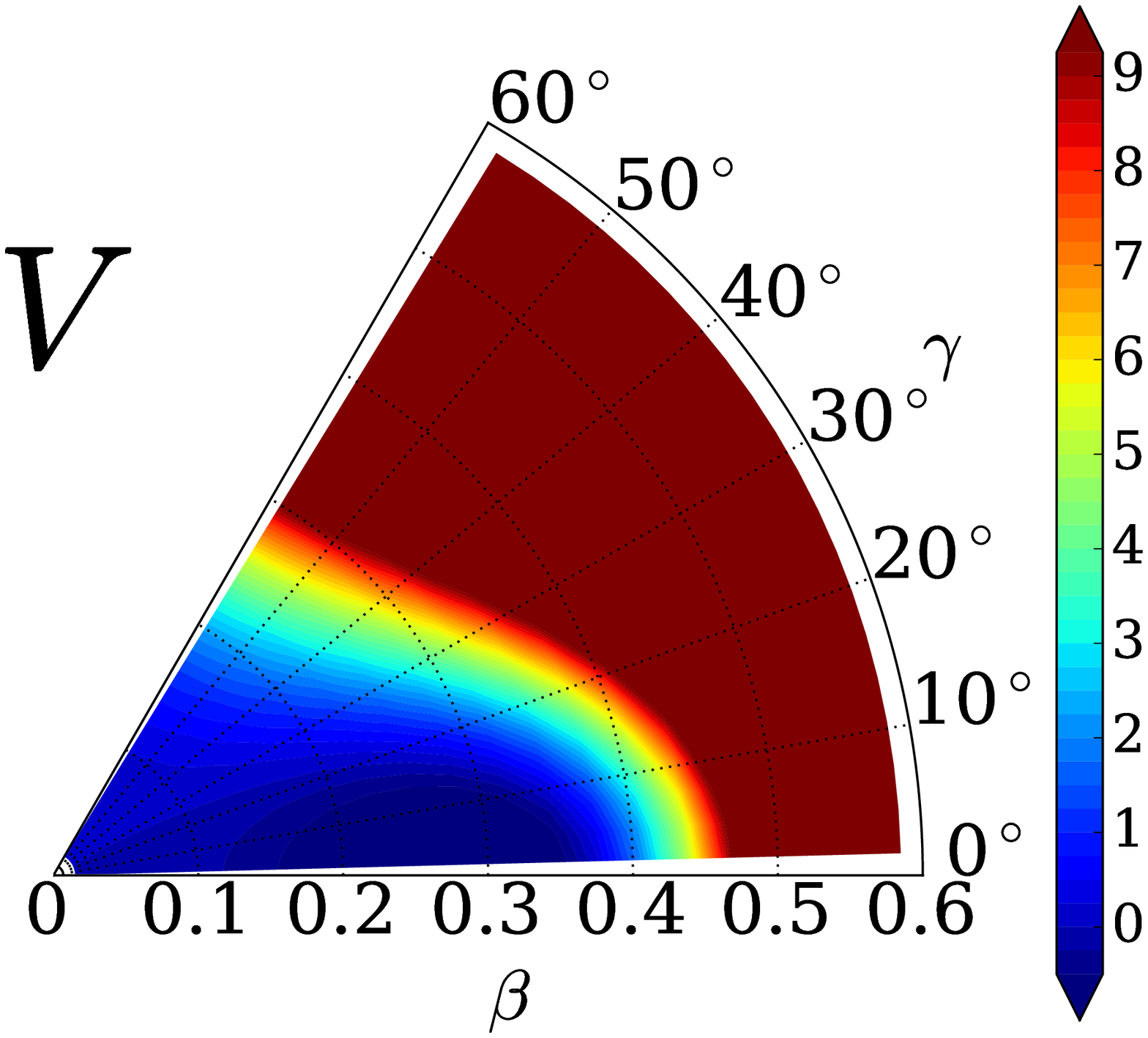}
  \vspace{0.5em}
  \subcaption{spherical-prolate transitional}\label{V:sph-pro}
 \end{minipage}
\caption{Collective potential $V_{\rm coll}(\beta, \gamma)$ [MeV] defined by Eq. (\ref{eq:collective potential}) with the parameter sets
 in Table 1.}\label{fig:V}
\end{figure}

The parameter sets in the collective potential and \KK{inertial} masses 
to simulate the three situations
are listed in Table 1.
In all the calculations, we set $D=50$ MeV$^{-1}$.
The spherical case is treated as the 5D harmonic oscillator (HO).  
We plot in Fig. \ref{fig:V} the collective potentials $V_{\rm coll}(\beta,\gamma)$
obtained with these three parameter sets.
While in the prolate case there is an absolute minimum around
$\beta=0.35$, the collective potential is soft along the $\beta$ direction
in the transitional case. The parameter $\epsilon^\prime$ in
Eqs. (\ref{eq:Dbb})--(\ref{eq:Dk}) controls the
oblate-prolate asymmetry. When $\epsilon^\prime$ is positive (negative), oblate
(prolate) shape is favored to reduce the rotational energy.
For these three parameter sets, we solve the collective Schr\"odinger
equation and obtain the excitation energies and collective wave
functions.
Note that these parameters are not adjusted by fitting to specific
experimental data but determined to simulate typical situations widely observed in the
nuclear chart. 
By scaling the parameters $C_i~(i=2, 4, 6), v_1,$ and $D^{-1}$
simultaneously,
one can scale the excitation energies with the collective wave functions
unchanged.
Therefore, only the ratios of the excitation energies between
excited states such as $R_{4/2}=E(4_1^+)/E(2_1^+)$ are meaningful.
The obtained $R_{4/2}$ values are 2.0, 2.2, \TF{and} 3.0 for the
spherical, transitional \TF{and} prolate parameter sets, respectively.
\SK{
Note that although this ``prolate'' parameter set simulates a prolately
deformed rotor, it is not an ideal rigid rotor but contains shape fluctuation.}
\KS{We have}
\TF{confirmed that the deviation between the ideal rigid rotor and prolately deformed rotor is very small on the proton elastic and inelastic scatterings.}

\begin{table}[tb]
\begin{center}
  \begin{tabular}{ll} \hline
                   & Choice of parameters\\ \hline \hline
spherical (5DHO)   & $C_2=50$ \\
prolately deformed & $C_4=800, \beta_0^2=0.1, v_1 =-150, C_6=1000,\epsilon=- 0.5$   \\ 
spherical-prolate transitional& $C_4=800, v_1 =-200, C_6=1000$  \\ 
\hline
  \end{tabular}
\end{center}
\caption{The parameter sets to simulate the three typical situations.
We take $D=50$ MeV$^{-1}$ for all the three calculations.
The parameters in Eqs. (\ref{eq:collective potential})--(\ref{eq:Dk}) 
which are not shown above are zero.
Here, $\epsilon$ is defined by $\epsilon^\prime= \epsilon/\beta_0$.
All the parameters other than the two dimensionless ones, $\beta_0$ and
 $\epsilon$, are in units of MeV.}
\label{table:parameters in the limiting cases}
\end{table}

For the eigenstates $\Psi_{\alpha IM}(\beta, \gamma, \Omega)$ obtained with the collective Hamiltonian for each
parameter set, 
we calculate the transition density matrix elements,
\begin{align}
\langle \alpha^\prime I^\prime M^\prime |\hat \rho_{n(p)}({\bm r})|\alpha I M\rangle
=\sum_{\mu\nu}
(IM\mu\nu|I\mu I^\prime M^\prime)
\rho_{n(p),\alpha^\prime I^\prime\alpha I}^{\mu}(r)Y_{\mu\nu}^*(\theta, \varphi), \label{eq-tr}
\end{align}
where the neutron (proton) transition density operator is given by
$\hat \rho_{n(p)} ({\bm r})= \sum_{i=1}^{N(Z)} \delta( {\bm r- \bm
r_i})$, and we have used the Wigner--Eckart theorem~\cite{Edmonds1996}.

To evaluate the above matrix elements,
the proton and neutron densities are needed 
for intrinsic states $\Phi_{\alpha I K}(\beta, \gamma)$.
As we are taking a phenomenological approach, 
we solve the deformed Woods--Saxon (WS) potential problem \KK{with
the $\beta$-$\gamma$ constraint}
on the mesh
point $(\beta_i, \gamma_j)$
instead of performing microscopic calculation \KK{such as constrained
Hartree--Fock--Bogoliubov calculation}.
Here we employ the following mesh points on the $\beta$-$\gamma$ plane:  
$\beta_i=(i-1/2)\Delta \beta, \gamma_i=(i-1/2)\Delta \gamma
~(i=1,\cdots,20)$ with $\Delta\beta=0.03$ and $\Delta\gamma=3.0^\circ$. 
For this calculation, we used TRIAXIAL2014~\cite{Mohammed-Azizi2014},
\KS{which solves one-body problem with the deformed WS, the spin-orbit, and
the Coulomb potentials.}
For the parameters of the deformed WS potential, 
the universal parameter set in Ref. \cite{Kahane1989} was adopted.

With the transition density, we obtain 
the proton elastic and inelastic cross sections based on the microscopic coupled-channel (MCC) calculation.
Namely, the diagonal and transition potentials used in the coupled-channel (CC) calculation are derived from the folding procedure.
In this Letter, we apply the JLM complex nucleon-nucleon
interaction~\cite{JLM1977} to the MCC calculation in the same as \SK{in} Ref.~\cite{takashina2008}.
The JLM interaction is usually written in the form \SK{of} 
\begin{equation}
v_{NN} (s; \rho, E) = \frac{V(\rho, E)}{(t \sqrt{\pi})^3} \exp{\left( -\frac{s^2}{t^2} \right)}
              + i \frac{W(\rho, E)}{(t \sqrt{\pi})^3} \exp{\left( -\frac{s^2}{t^2} \right)},
\end{equation}
where $V(\rho, E)$ and $W(\rho, E)$ are the strength of the real and imaginary parts, respectively.
They include the isoscalar and isovector components.
$\rho$ and $E$ are the nucleon density and the incident proton energy, respectively.
\TT{$t$ is the range parameter} of the nucleon-nucleon interaction. 
\TT{We fix the $t$ value to be 1.2 according to Ref.~\cite{JLM1977}.
The renormalization factor, which is often used to adjust the strength of the potential based on the folding model, is not applied in this Letter.}

We apply our model introduced above to the \TF{proton elastic and inelastic scatterings by $^{154}$Sm.}
It is well known that a spherical-to-prolate shape transition occurs 
with increasing the neutron number
in samarium isotopes in this mass region. 
Actually, the experimental value of $R_{4/2}$ for $^{154}$Sm is 3.2 and
the experimental $\beta_2$ is 0.34~\cite{NNDC}, which implies
that the shape phase transition to the deformed shape already occurred.
Although $^{150}$Sm exhibits more transitional character ($R_{4/2}^{\rm
exp.}= 2.3$), we \TF{will show the results of} 
$^{154}$Sm for the following reasons.
First, 
\TF{the abundance of $^{154}$Sm is enough and the experimental data is also plenty.}
Second, in this \KK{phenomenological} analysis the difference in the neutron number by four only
plays a minor role.

\begin{figure}[tb]
\centering
  \includegraphics[keepaspectratio, scale=0.35,clip]{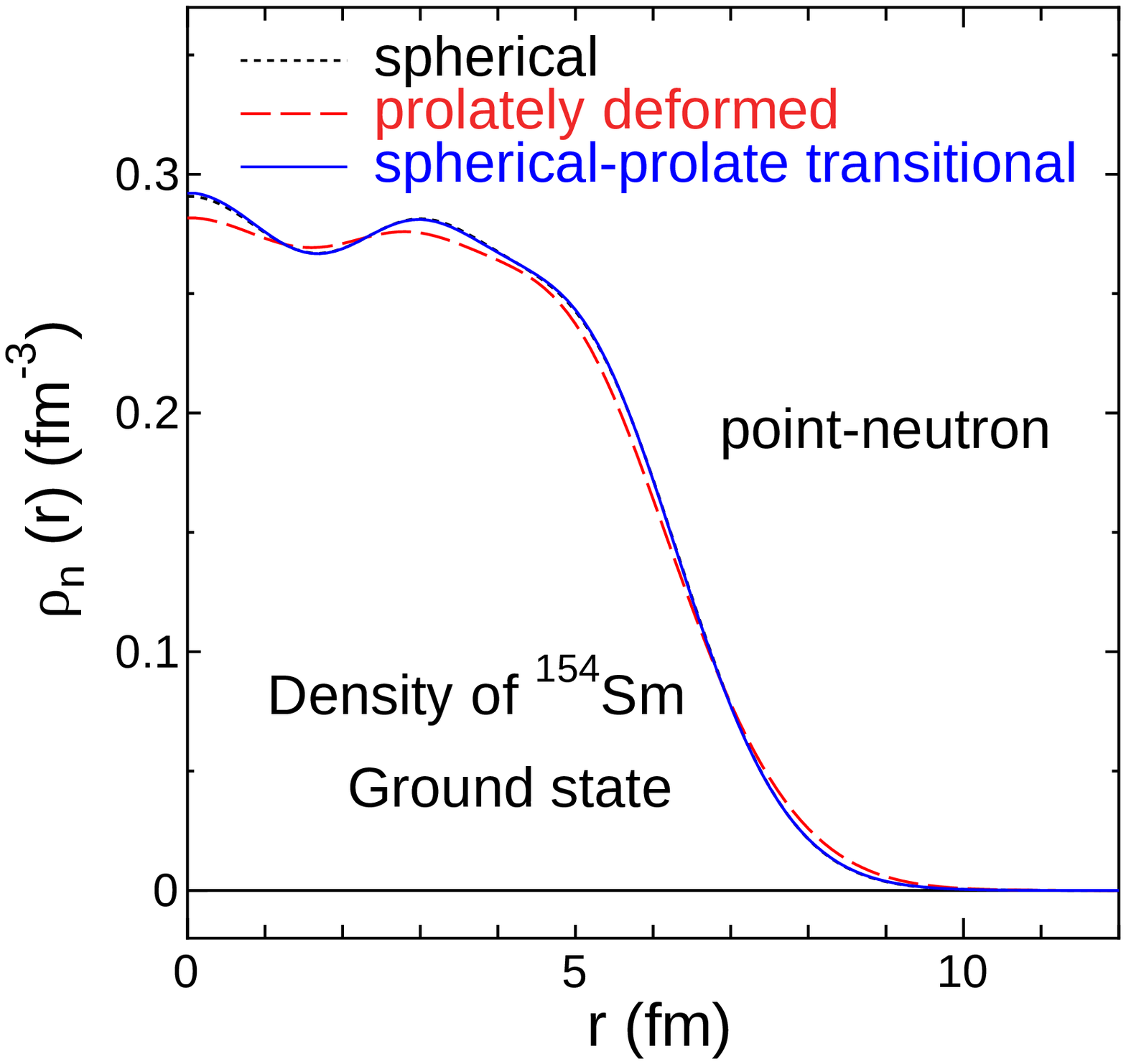} 
  \includegraphics[keepaspectratio, scale=0.35,clip]{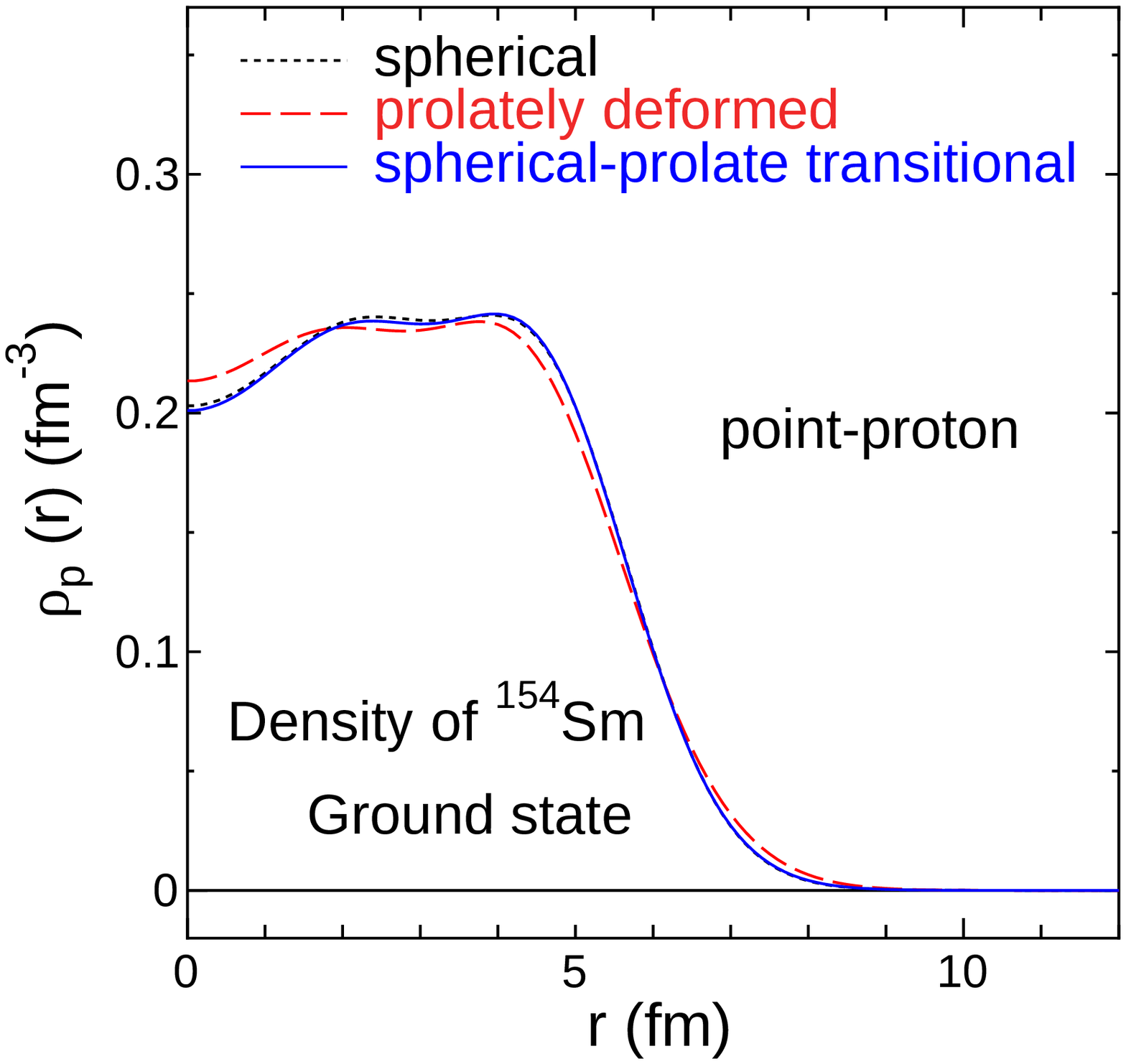} 
\caption{The point-neutron and point-proton \SK{density} distributions
 obtained with the spherical (dotted line), prolate (dashed), and
 transitional (solid) models.}
\label{density_Sm154}
\end{figure}

\SK{Figure} \ref{density_Sm154} shows the point-neutron and point-proton densities distributions \TT{for the ground state}
obtained with the spherical, prolate, and transitional models.
The neutron and proton density distributions in the transitional model almost coincide
with those in the spherical HO model.
\KK{We can see a difference between} the density distribution derived from the prolately deformed model 
\KK{and those derived from} the spherical HO and transitional models, especially for the tail part.
\KK{In the prolate case,} the tail part of the nuclear density \TT{in the
laboratory frame} is expanded by the deformation.
\KK{The reason why the spherical and transitional models give similar
density distributions can be attributed to their
ground-state collective wave functions.
The ground-state collective wave function in the transitional model [shown in Fig. 6(a)]
spreads around the sphericity and is similar to that in the spherical
HO model (not shown here), which leads to almost the same density
distribution in the ground state.}
In Ref.~\cite{minomo2011}, the deformation effect is discussed on the total reaction cross section.
\KK{Below}, the effect of the difference in the density distributions
will be \KK{briefly} discussed on the proton elastic cross section.

\begin{figure}[tbp]
 \begin{minipage}[b]{0.33\linewidth}
  \centering
  \includegraphics[keepaspectratio,  scale=0.28,clip]{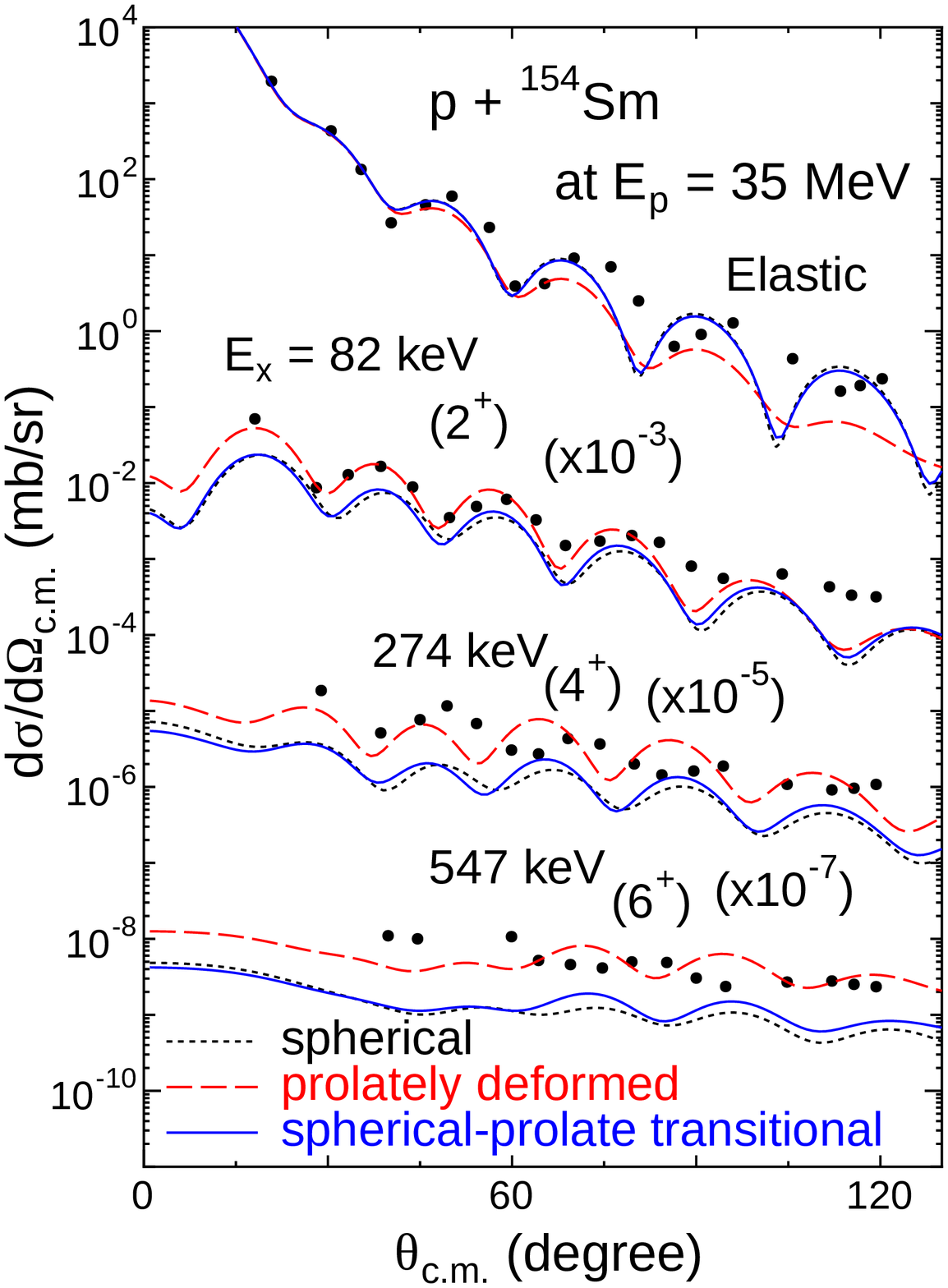}
  \vspace{0.5em}
  \subcaption{$E_p=35$MeV}\label{xs_Sm154_0350_el+inl}
 \end{minipage}
\hspace{-1.25em}
 \begin{minipage}[b]{0.33\linewidth}
  \centering
  \includegraphics[keepaspectratio, scale=0.28,clip]{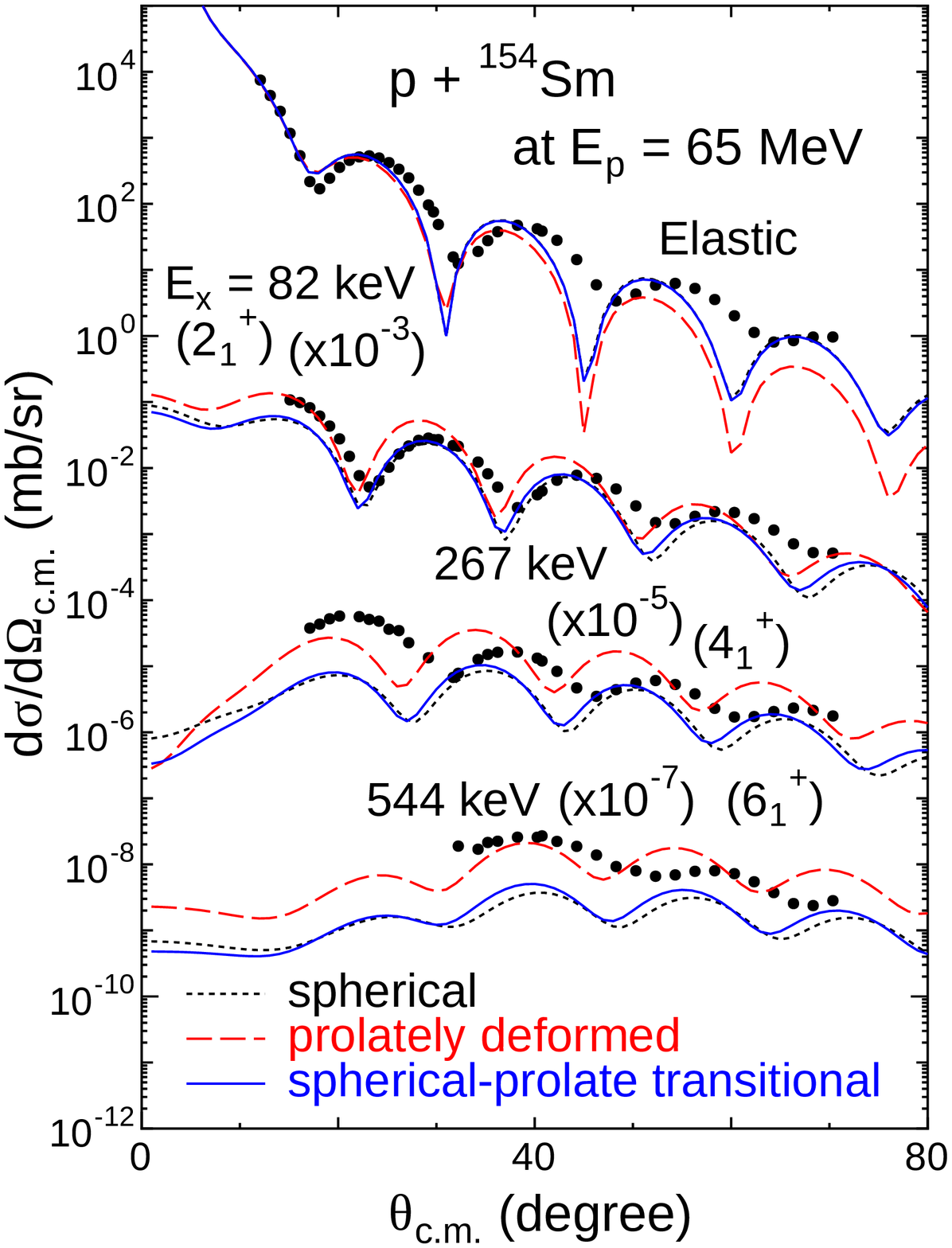}
  \vspace{0.5em}
  \subcaption{$E_p=65$MeV}\label{xs_Sm154_0650_el+inl}
 \end{minipage}
\hspace{-1.25em}
 \begin{minipage}[b]{0.33\linewidth}
  \centering
  \includegraphics[keepaspectratio, scale=0.28,clip]{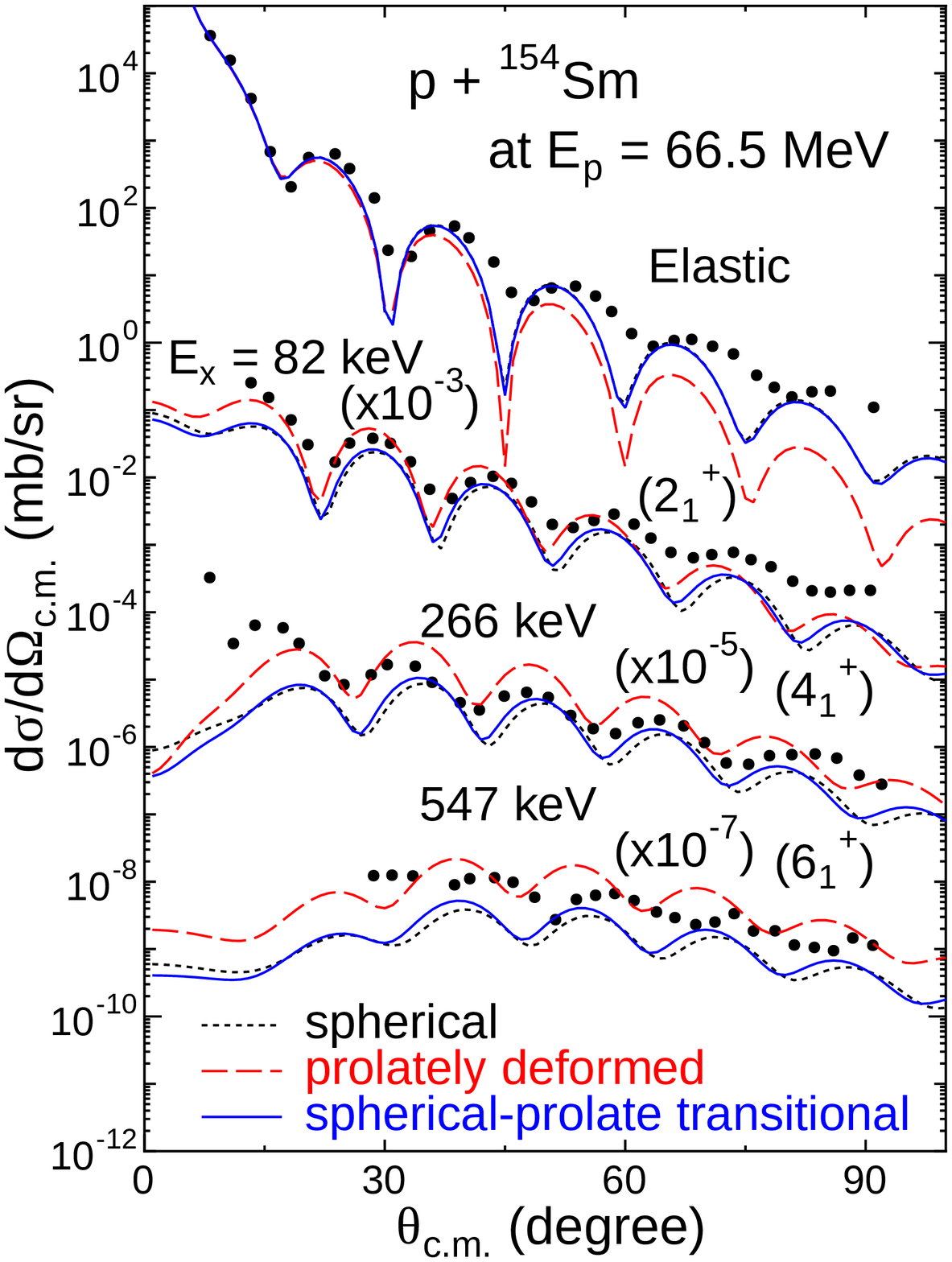}
  \vspace{0.5em}
  \subcaption{$E_p=66.5$MeV}\label{xs_Sm154_0665_el+inl}
 \end{minipage}
\caption{The calculated elastic and inelastic differential cross sections \TT{of the $p$ + $^{154}$Sm system}
in comparison with experimental
 data~\cite{exp-0350, exp-0650, exp-0665} \TT{at} $E_p=$35, 65, \TF{and
 66.5} MeV.
The differential cross sections for the spherical, prolate, and
 transitional collective Hamiltonians are indicated by dotted, dashed, 
and solid lines, respectively.
}\label{fig:xs_Sm154_el+inl}
\end{figure}

We show in Fig. \ref{fig:xs_Sm154_el+inl} the calculated  
elastic and inelastic \TF{scattering}  cross sections 
at $E_p=35$, 65, \TF{and 66.5} MeV in comparison with the experimental data.
For the elastic scattering, 
the calculated cross sections 
\SK{reasonably} reproduce the experimental data. 
\TT{In detail,} 
\KK{the calculated results for the prolate model deviate from those for the 
spherical and transitional models for backward angle, which is}
caused by the difference in the tail part of the density distribution of the ground state as mentioned above.

\KK{Note that, in this calculation, we have not adjusted any parameters
in our model (those in the deformed
WS model, the model collective Hamiltonians, and the JLM interaction), 
other than a rough adjustment of $\beta_0=0.33$ for the prolate rotor model.
We have seen that the backward elastic scattering may be sensitive to 
the tail part of the ground-state density distribution.
It can be affected not only by deformation but also by the diffuseness of
the nuclear surface in the intrinsic frame. 
We have not adjusted the diffuseness parameters in the deformed WS potential
as mentioned above.       
}

For the inelastic scattering, 
\TF{the calculated cross sections also reproduce the experimental data.}
There is little difference between the results obtained with the spherical HO and the
transitional models \TF{not only for the elastic differential cross sections but also for the}
inelastic differential cross sections for the yrast
$2^+, 4^+,$ and $6^+$ states.

\begin{figure}[tb]
 \begin{minipage}[b]{0.33\linewidth}
  \centering
  \includegraphics[keepaspectratio,  scale=0.28,clip]{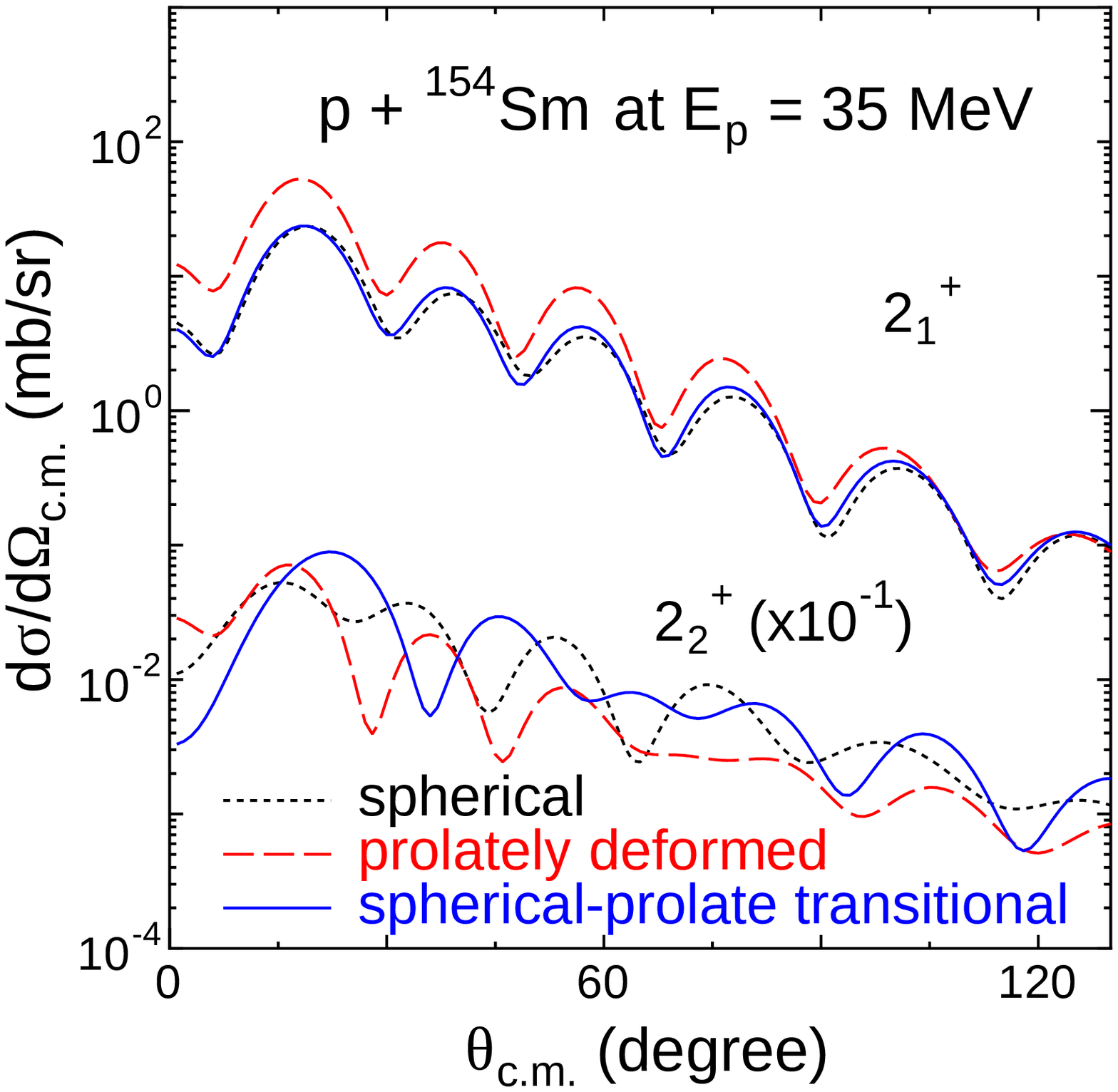}
  \vspace{0.5em}
  \subcaption{$E_p=35$MeV}\label{xs_Sm154_0350_2plus}
 \end{minipage}
\hspace{-1.25em}
 \begin{minipage}[b]{0.33\linewidth}
  \centering
  \includegraphics[keepaspectratio, scale=0.28,clip]{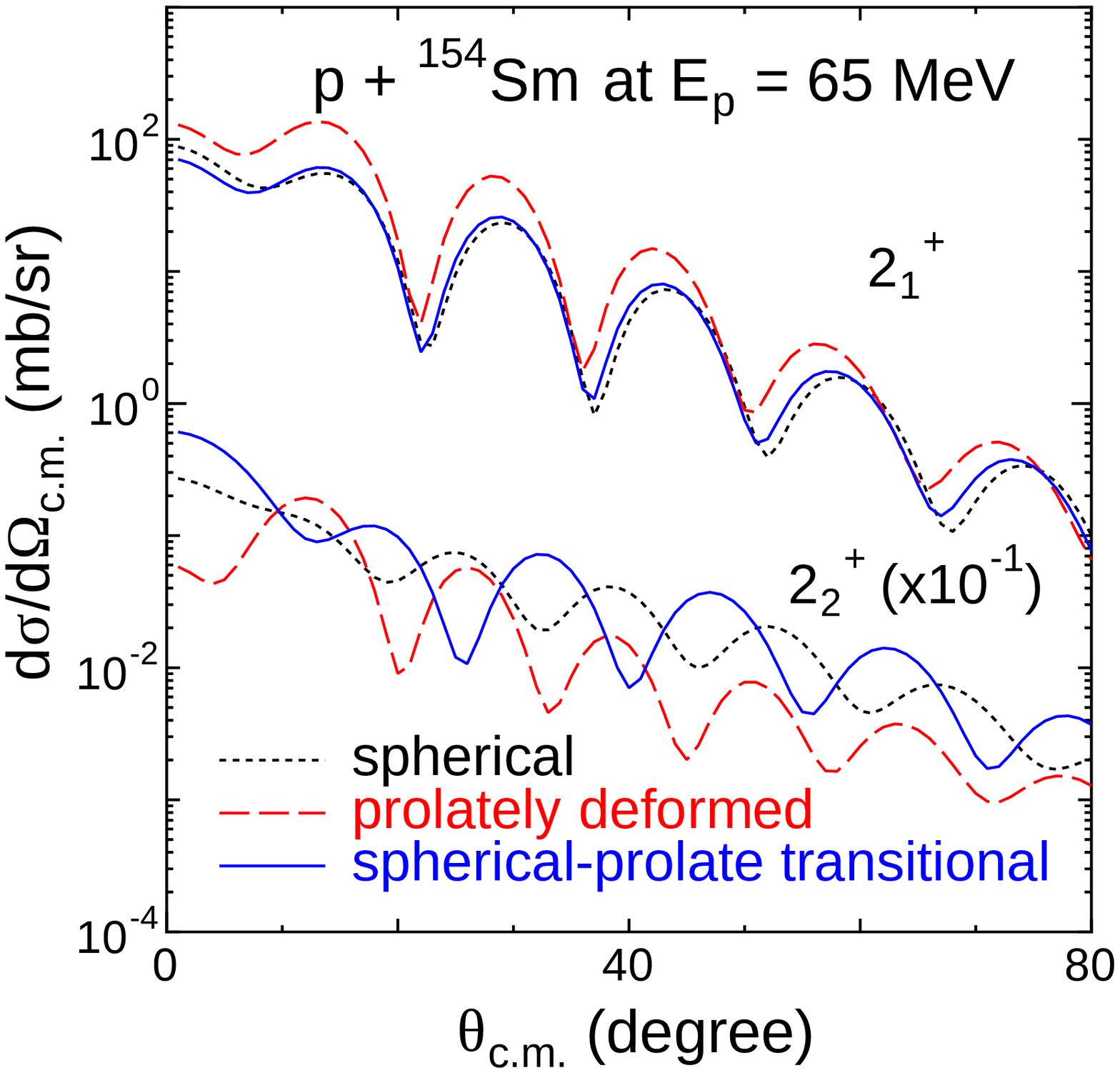}
  \vspace{0.5em}
  \subcaption{$E_p=65$MeV}\label{xs_Sm154_0650_2plus}
 \end{minipage}
\hspace{-1.25em}
 \begin{minipage}[b]{0.33\linewidth}
  \centering
  \includegraphics[keepaspectratio, scale=0.28,clip]{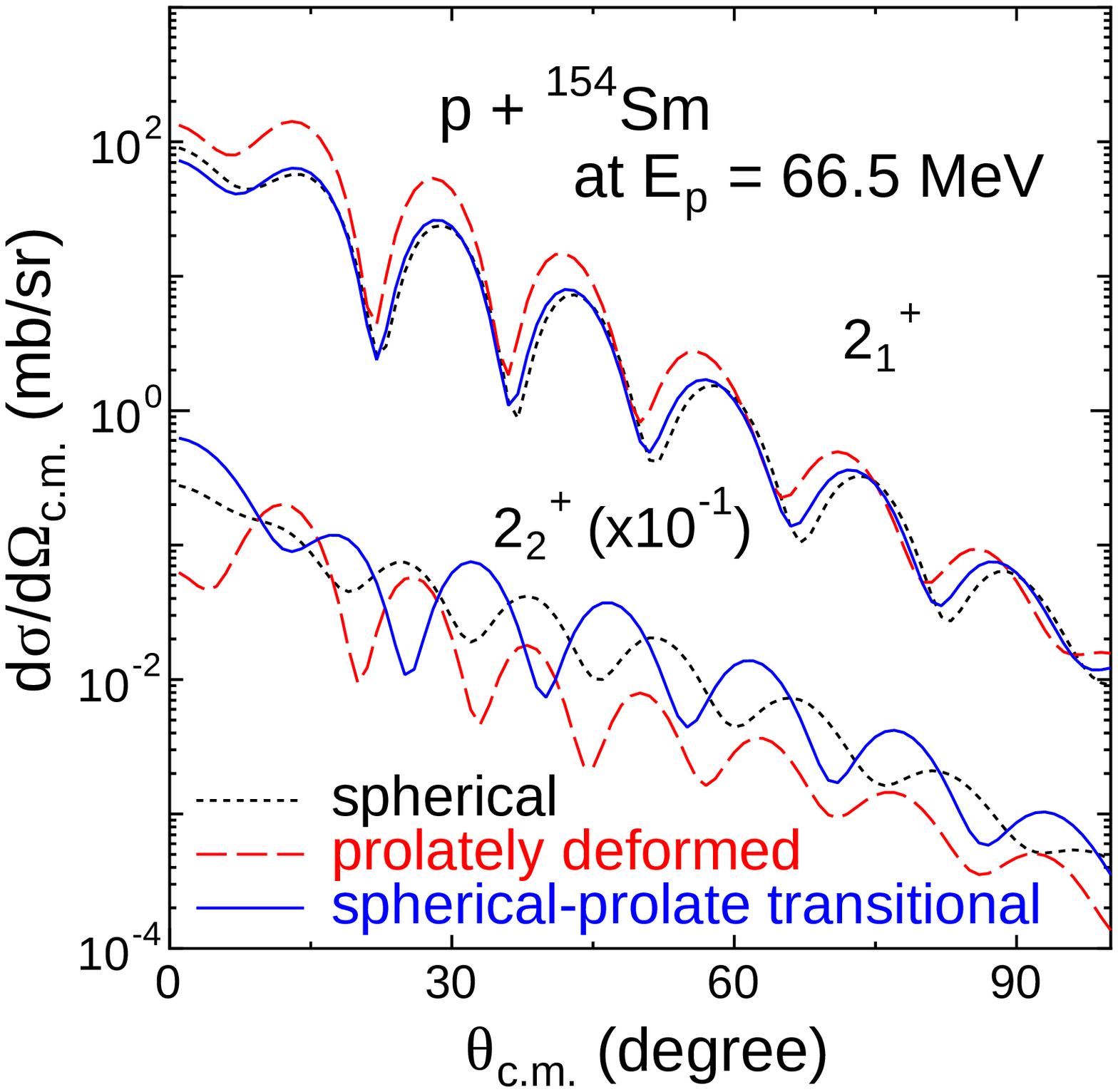}
  \vspace{0.5em}
  \subcaption{$E_p=66.5$MeV}\label{xs_Sm154_0665_2plus}
 \end{minipage}
\caption{
Same as Fig.~\ref{fig:xs_Sm154_el+inl} but for the inelastic
 differential cross section for  the $2^+_1$ and $2_2^+$ states.
}
\label{xs_0650_Sm154_inl-2plus}
\end{figure}

\TF{We apply our models to the non-yrast states to investigate the effect of the quadrupole shape mixing.}
In Fig. \ref{xs_0650_Sm154_inl-2plus}, the inelastic differential cross
sections at $E_p=$ \TF{35, 65, and 66.5} MeV 
for the \TT{ 
$2_2^+$}  state are displayed.
Here, we plot the inelastic differential cross sections for the $2_1^+$
state again for comparison of the diffraction pattern.
\TT{For the $2_1^+$ state, the \KK{three models give} similar angular distributions.
On the other hand, the calculated diffraction patterns are completely different for the $2_2^+$ state.
Especially, }
whereas 
the positions of the peaks for the spherical and prolate
\KK{models} are almost the same,
we clearly observe that
the positions of peaks for the transitional model
are shifted to backward compared with those of the other two models.
\SK{This shift turns out to remain even if we disregard the multistep processes.
Thus, it will be due to the difference in the calculated transition densities.}

\begin{figure}[tbp]
\centering
 \begin{minipage}[b]{0.5\linewidth}
  \centering
  \includegraphics[keepaspectratio, scale=0.28,clip]{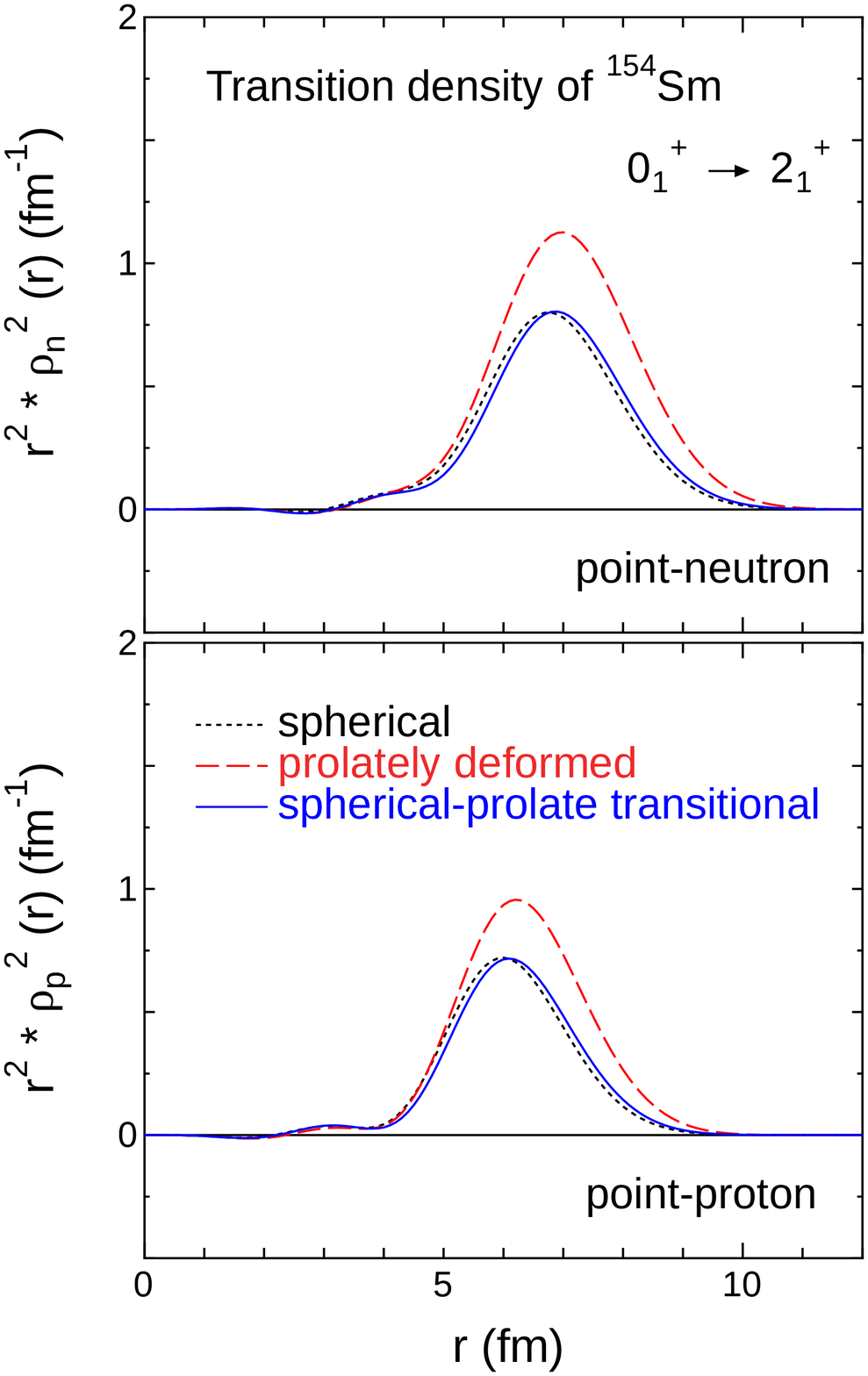}
  \vspace{0.5em}
  \subcaption{$0_1^+ \rightarrow 2_1^+$}\label{tr-density-r2_154Sm_gs-1st2plus}
 \end{minipage}
\hspace{-1.25em}
 \begin{minipage}[b]{0.5\linewidth}
  \centering
  \includegraphics[keepaspectratio, scale=0.28,clip]{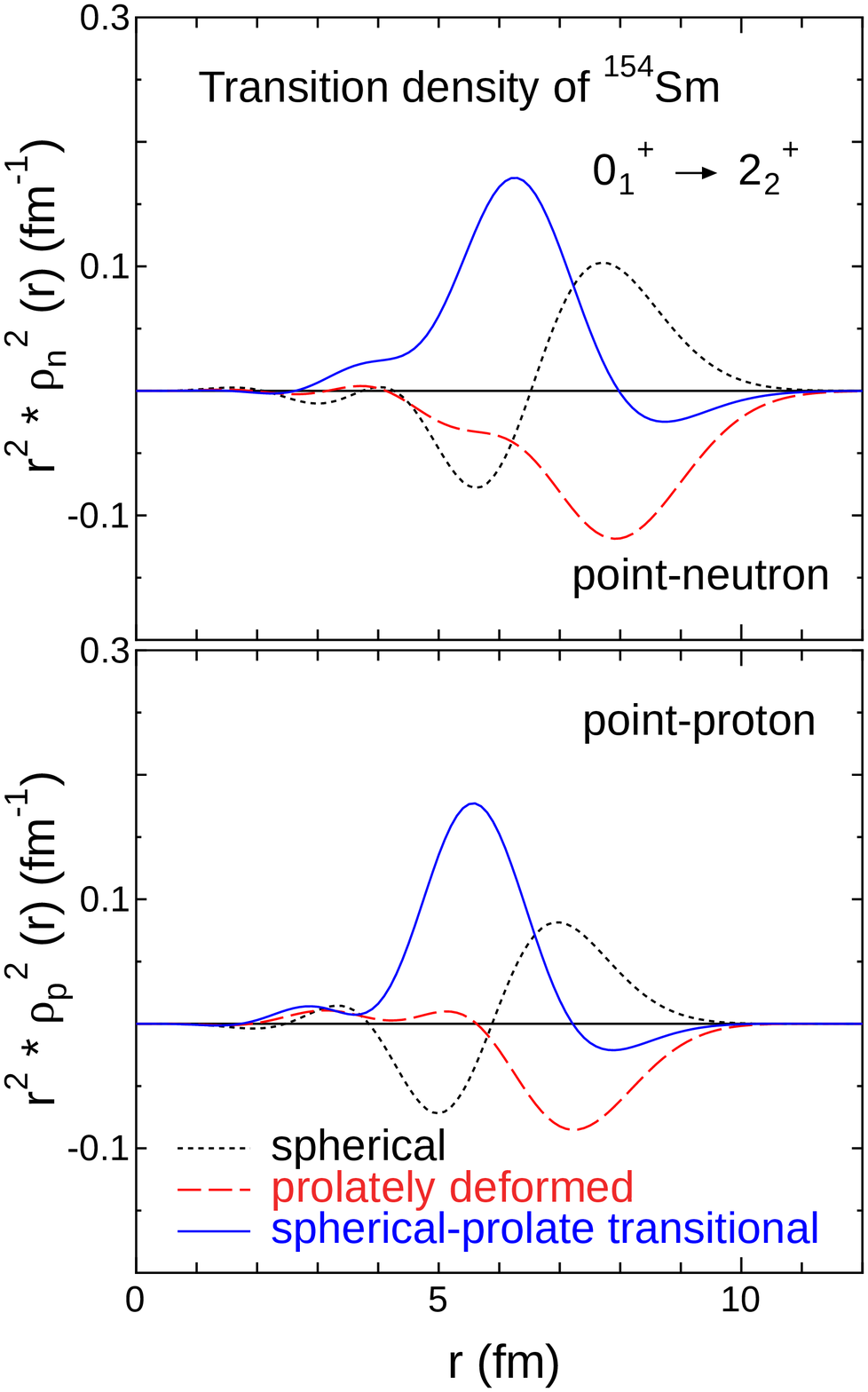}
  \vspace{0.5em}
  \subcaption{$0_1^+ \rightarrow 2_2^+$}\label{tr-density-r2_154Sm_gs-2nd2plus}
 \end{minipage}
\caption{The neutron (proton) transition densities multiplied by $r^2$, $r^2\rho^{2}_{n(p)}(r)$, from the ground state to the first and
 second $2^+$ states in $^{154}$Sm. The calculated results for the
 spherical, prolate, and transitional parameter sets are indicated by
 dotted, dashed, and solid lines, respectively.}\label{tr-density-r2_154Sm}
\end{figure}

We plot in Fig. \ref{tr-density-r2_154Sm} the transition densities
multiplied by $r^2$, $r^2\rho^{2}_{n(p)}(r)$,
for the $0_1^+ \rightarrow 2_1^+$ and $0_1^+ \rightarrow 2_2^+$
transitions.
For the $0_1^+ \rightarrow 2_1^+$ transition, 
although the peak height for the prolate shape is larger than the
other two, the structures of the transition densities obtained with
the three calculations are similar to one another.
On the other hand, the transition densities for
the $0_1^+ \rightarrow 2_2^+$ transition exhibit a rather different behavior.
Because the neutron and proton transition densities
have similar structure, we shall focus on the neutron transition density
below.
We see that, in the spherical and prolate cases, the main peak of the transition density  
is located around $r=8$ fm.
In the transitional case, the main peak is located in an inner region
around $r=6$ fm.
%

\begin{figure}[h]
 \begin{minipage}[b]{0.32\linewidth}
  \centering
  \includegraphics[keepaspectratio, scale=0.2,trim=15 0 0 0,clip]{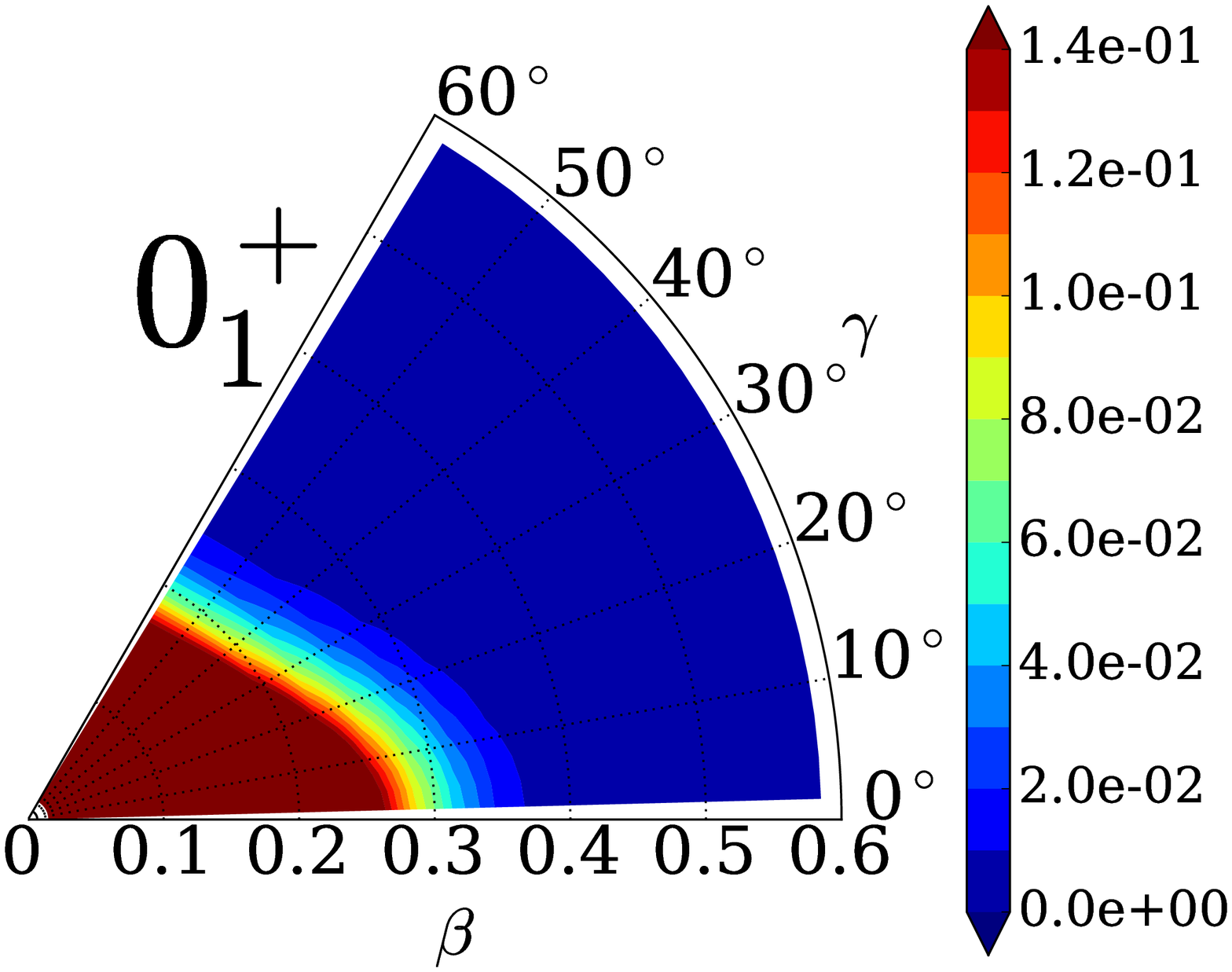}
  \vspace{0.5em}
  \subcaption{transitional $0_1^+$}\label{wave_sph-pro_C800_v1-200_C6_1000-I0k0}
 \end{minipage}
\hspace{-1.25em}
 \begin{minipage}[b]{0.32\linewidth}
  \centering
  \includegraphics[keepaspectratio, scale=0.2,trim=15 0 0 0,clip]{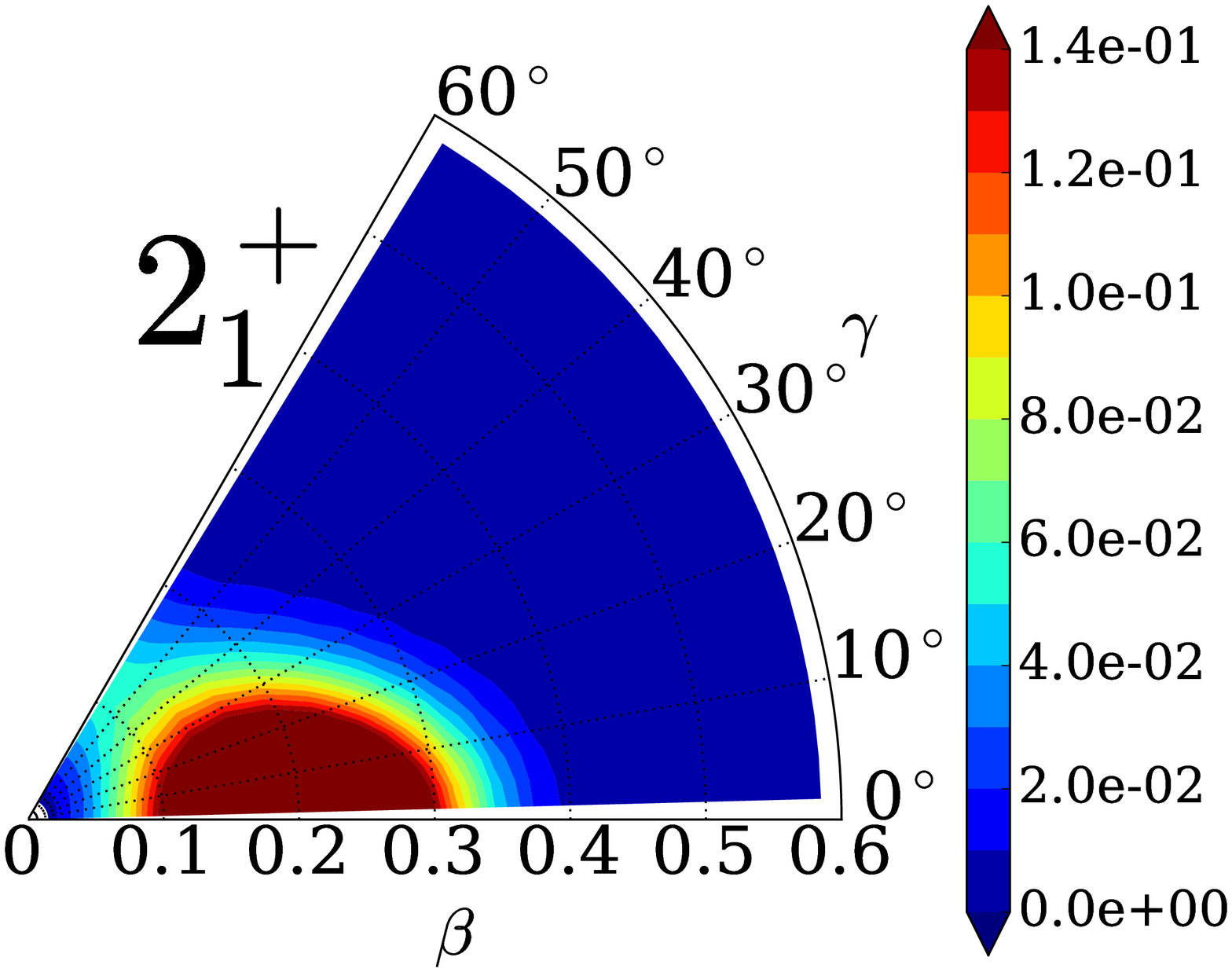}
  \vspace{0.5em}
  \subcaption{transitional $2_1^+$}\label{wave_sph-pro_C800_v1-200_C6_1000-I2k0}
 \end{minipage}
\hspace{-1.25em}
 \begin{minipage}[b]{0.32\linewidth}
  \centering
  \includegraphics[keepaspectratio, scale=0.2,trim=15 0 0 0,clip]{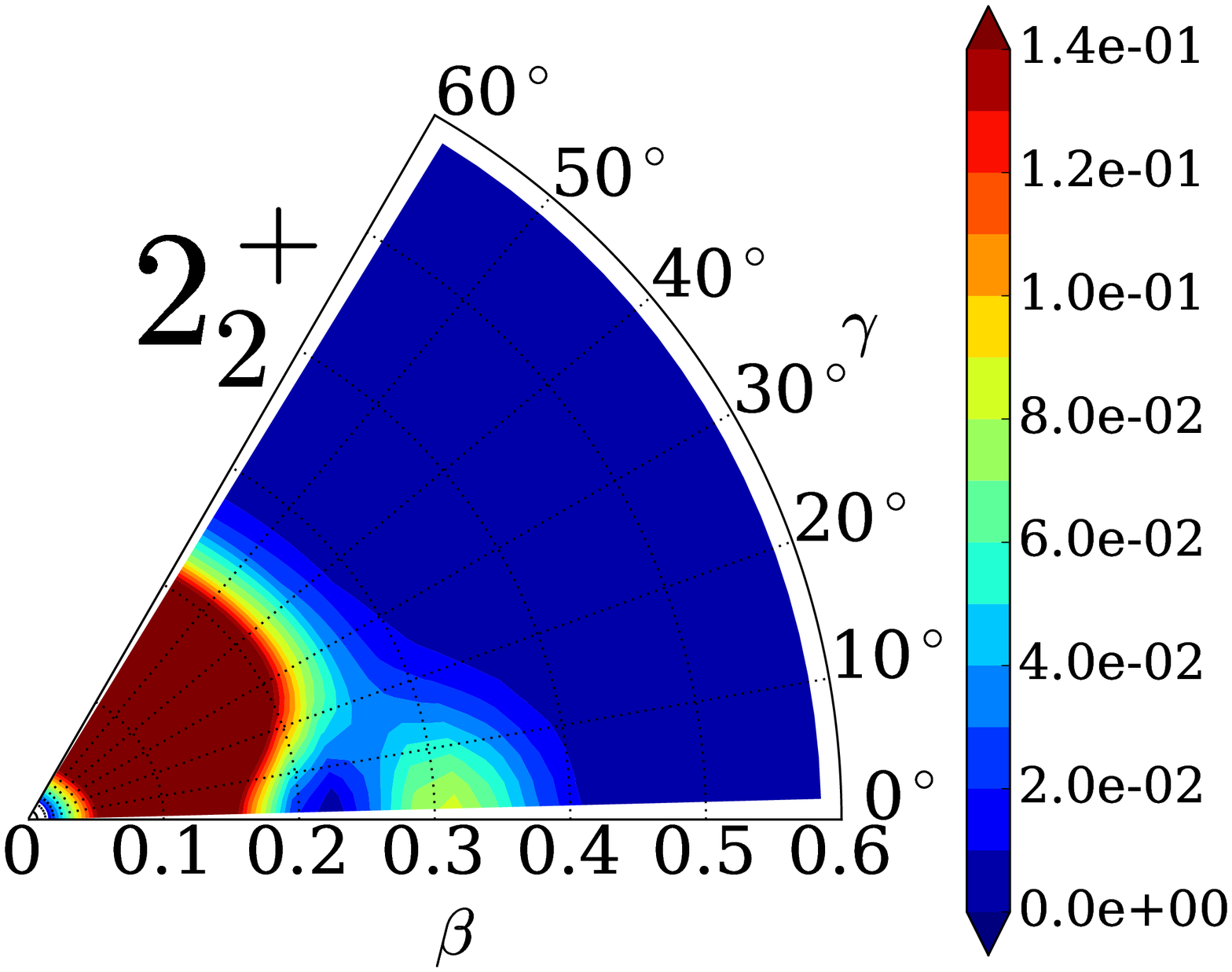}
  \vspace{0.5em}
  \subcaption{transitional $2_2^+$}\label{wave_sph-pro_C800_v1-200_C6_1000-I2k1}
 \end{minipage}
\\
 \begin{minipage}[b]{0.32\linewidth}
  \centering
  \includegraphics[keepaspectratio, scale=0.2,trim=15 0 0 0,clip]{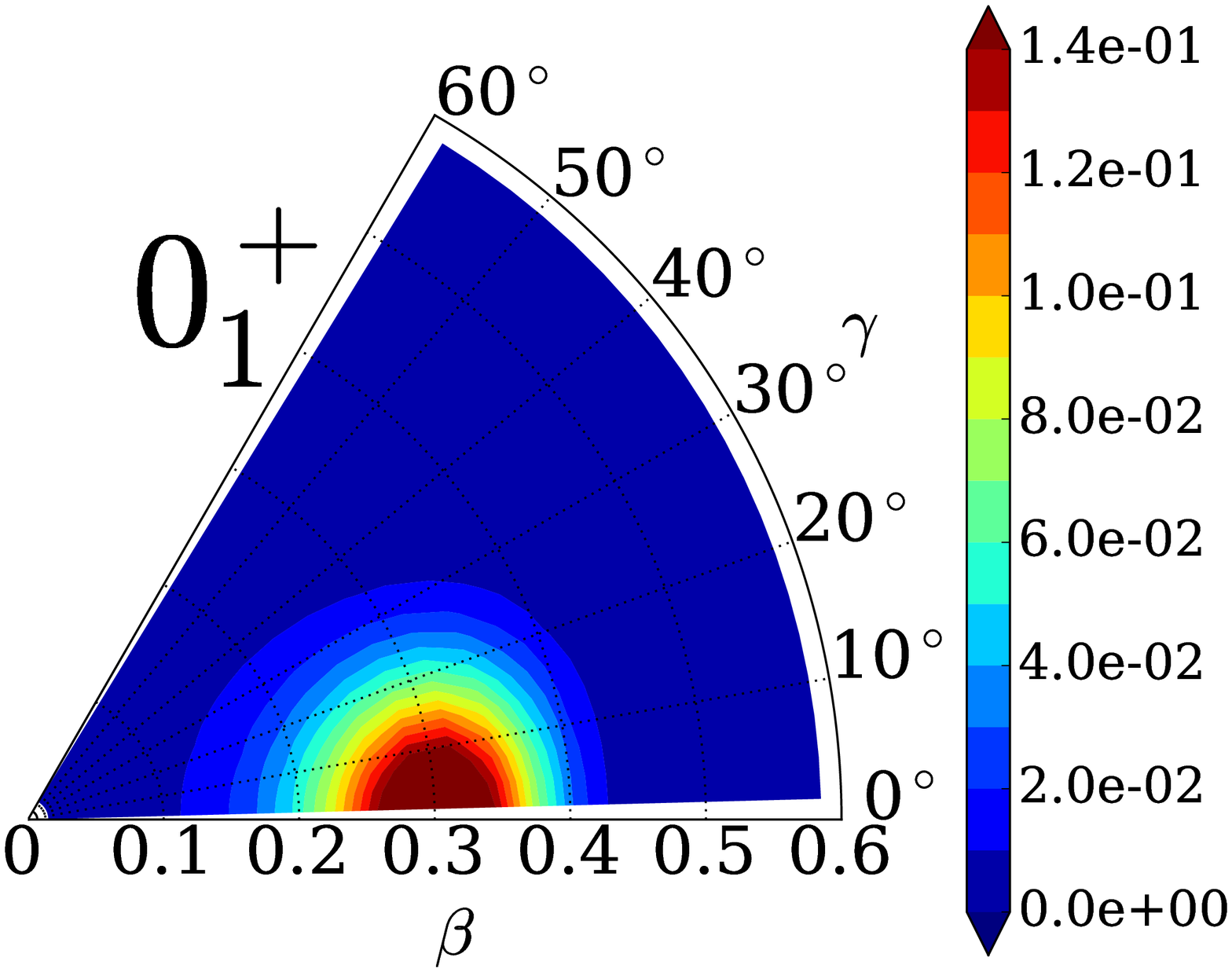}
  \vspace{0.5em}
  \subcaption{prolate $0_1^+$}\label{prolate_C800_b0sq0.1_v1-150_C6_1000_eps-0.5-I0k0}
 \end{minipage}
\hspace{-1.25em}
 \begin{minipage}[b]{0.32\linewidth}
  \centering
  \includegraphics[keepaspectratio, scale=0.2,trim=15 0 0 0,clip]{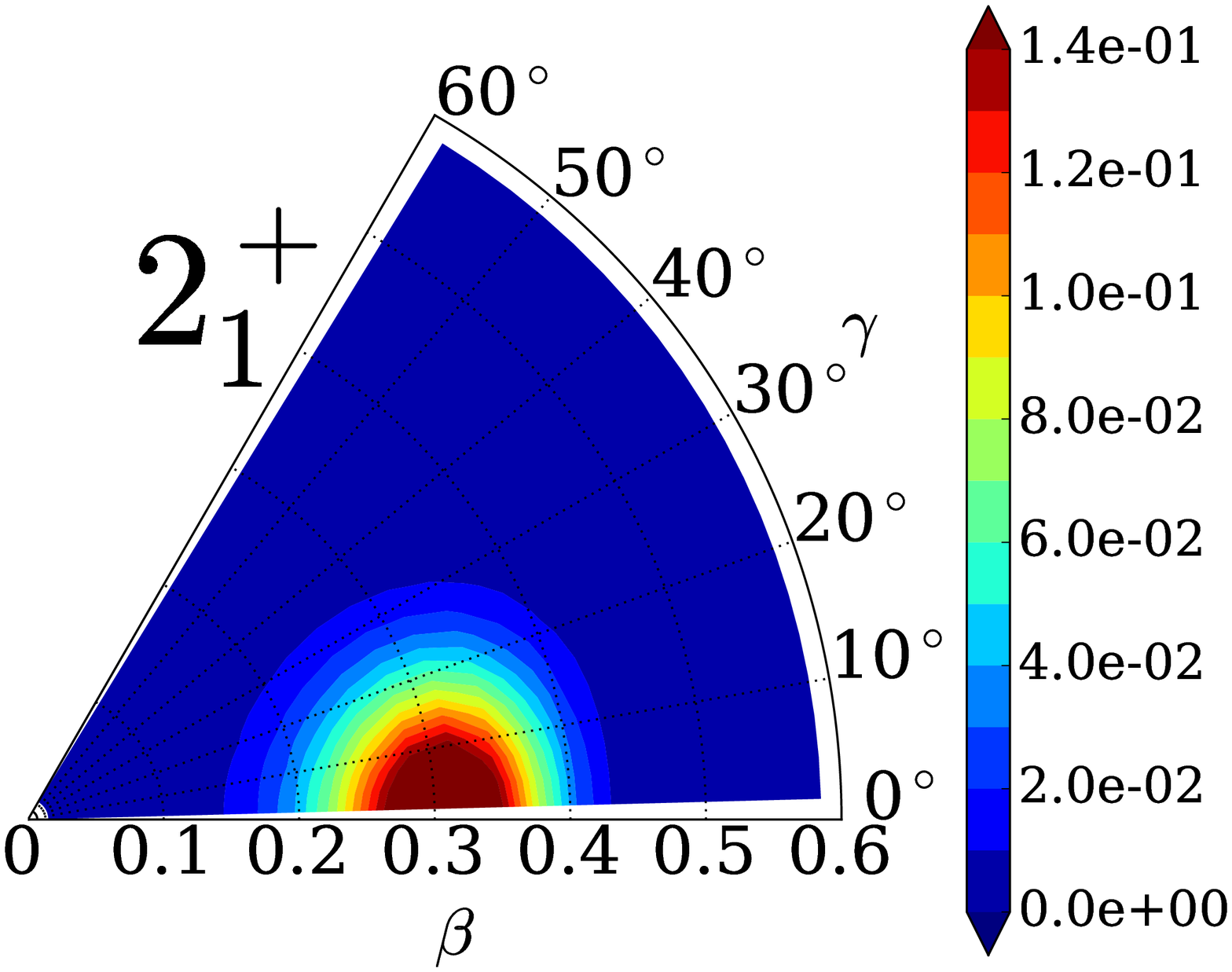}
  \vspace{0.5em}
  \subcaption{prolate $2_1^+$}\label{prolate_C800_b0sq0.1_v1-150_C6_1000_eps-0.5-I2k0}
 \end{minipage}
\hspace{-1.25em}
 \begin{minipage}[b]{0.32\linewidth}
  \centering
  \includegraphics[keepaspectratio, scale=0.2,trim=15 0 0 0,clip]{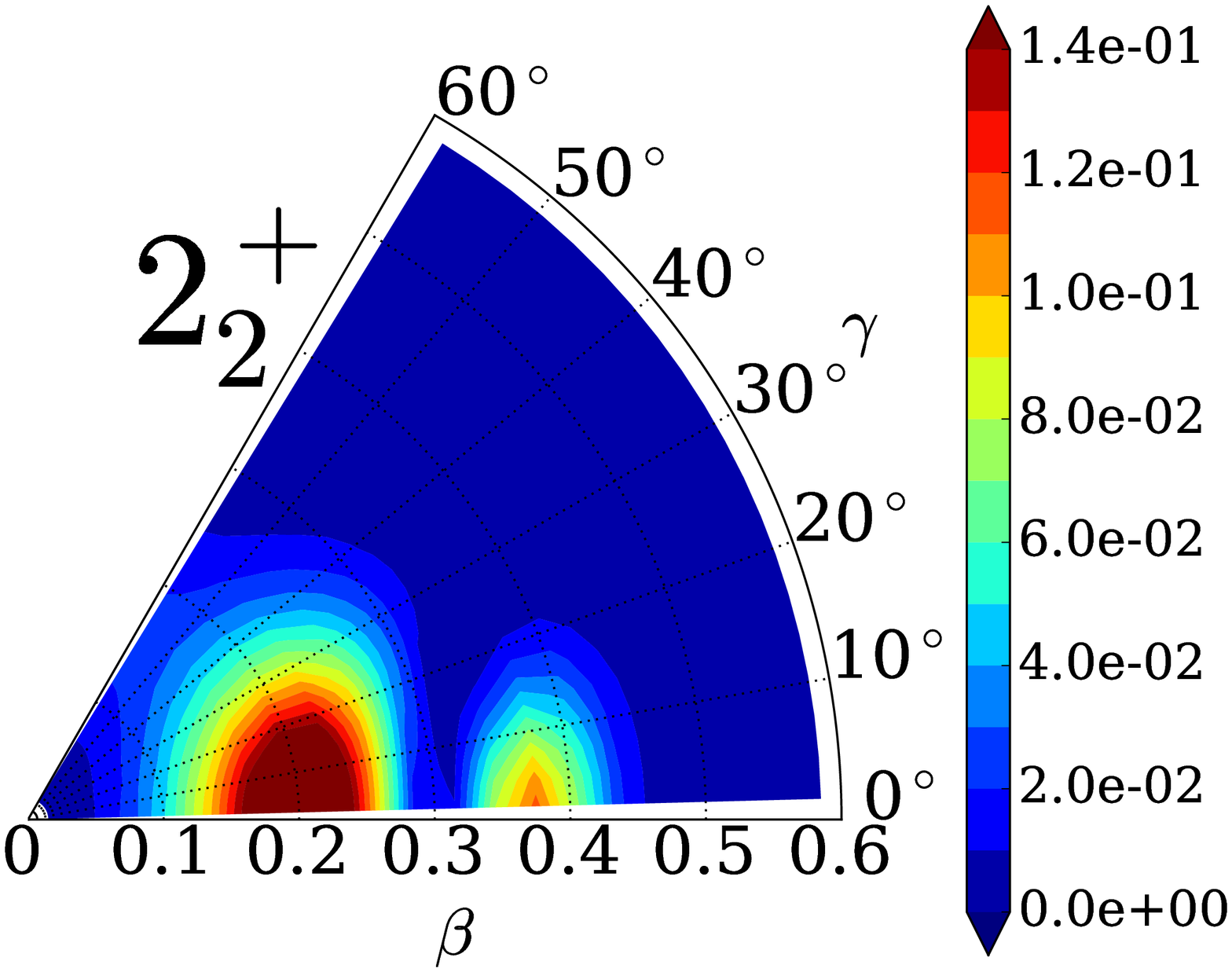}
  \vspace{0.5em}
  \subcaption{prolate $2_2^+$}\label{prolate_C800_b0sq0.1_v1-150_C6_1000_eps-0.5-I2k1}
 \end{minipage}
 \caption{The collective wave function squared $\sum_K |\Phi_{\alpha
 IK}(\beta, \gamma)|^2$ calculated for the $0_1^+, 2_1^+$, and $2_2^+$ states.}
\label{fig:collective wf}
\end{figure}

\begin{figure}[tbp]
\centering
 \begin{minipage}[b]{0.4\linewidth}
  \centering
  \includegraphics[keepaspectratio, scale=0.35,clip]{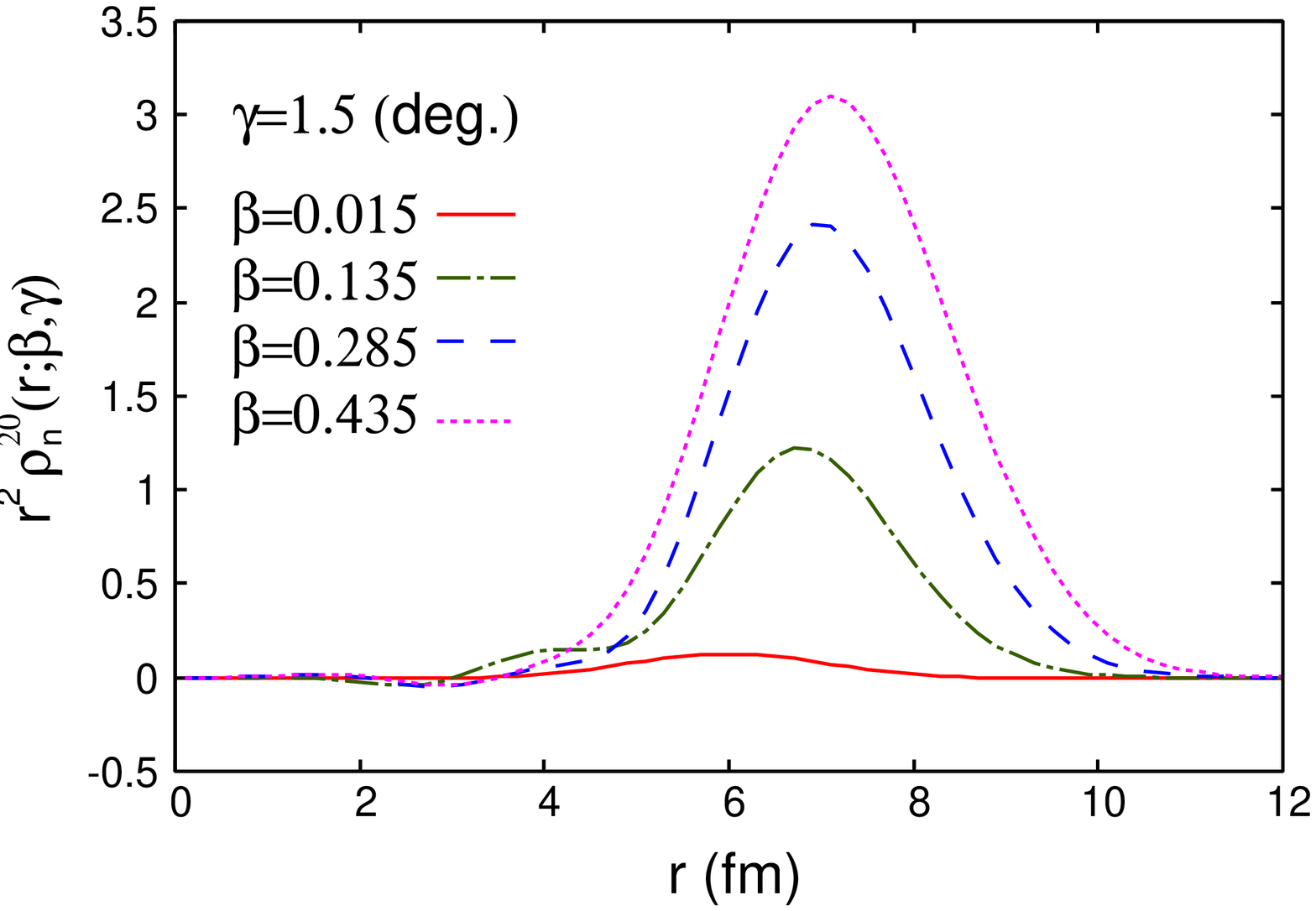}
  \vspace{0.5em}
 \end{minipage}
\caption{The $Y_{20}$ component of the neutron density \KK{multiplied by
 $r^2$}, $r^2\rho^{20}_{\rm n}(r;\beta, \gamma)$,
for the intrinsic states with $(\beta ,\gamma=1.5^\circ )$. }
\label{rho_20_r2_Z62N92_bxxg01}
\end{figure}

One may understand the difference between the prolate and transitional
cases in a relatively simple way as follows, while a more detailed
analysis is required for the spherical case.
We show in Fig. \ref{fig:collective wf} the collective wave functions squared 
$\sum_K |\Phi_{\alpha IK}(\beta, \gamma)|^2$ calculated for the $0_1^+, 2_1^+$, and $2_2^+$ states.
While, in the prolate case, 
the collective wave function squared in the ground band are localized around the prolate
potential minimum, that for the $2_2^+$ state has a two-peak structure on the prolate side. 
In this case, the $K=0$ component of the collective wave functions
dominates over the $K \neq 0$ components, and 
the $Y_{20}$ component of the intrinsic density gives a main
contribution to the transition density from $0_1^+$ to $2^+$ states.
We also plot in Fig. \ref{rho_20_r2_Z62N92_bxxg01} the $Y_{20}$
component of the neutron density $\rho^{20}(r;\beta, \gamma)$
for the intrinsic state with $(\beta ,\gamma=1.5^\circ )$.
One can see that the peak of $\rho^{20}_n(r;\beta, \gamma)$ develops with increasing $\beta$. 
The $2_2^+$ state is a $\beta$-vibrational state,
and the collective wave function $\Phi_{2_2^+, K=0}(\beta,\gamma)$ has a
node around $\beta=0.3$. 
\KK{The first peak around $\beta=0.2$ and the second peak around $\beta=0.4$
give a positive and negative contributions to the transition density, respectively.
The contribution from the second peak
dominates over the first}, which leads to the transition
density shown in Fig. \ref{tr-density-r2_154Sm}(b).

In the transitional case, 
we can see that, there is  strong $\beta$-$\gamma$ coupling, and 
the collective wave function of the $2_2^+$
state exhibits the $\gamma$-vibrational character as well as the
$\beta$-vibrational one.
\KK{There is a prolate peak around $\beta=0.3$ as in the prolate
case, although the peak height is smaller.
This peak contributes to a dip of the transition
density in an outer region.}  
The other component of the collective wave function squared spreading over the triaxial-oblate
region gives a dominant contribution to the
transition density shown in Fig. \ref{tr-density-r2_154Sm}(b).
Thus, the strong shape mixing in transitional nuclei may affect the inelastic 
differential cross sections.


In this study, 
\KK{
to discuss a possible observation of large-amplitude shape mixing in
transitional nuclei by nuclear reaction,
we adopted a phenomenological model 
based on the 5D quadrupole collective Hamiltonian
simulating isotopes exhibiting the spherical-to-prolate shape transition,
and  investigated the effect of the large-amplitude
quadrupole shape mixing on the \TF{proton elastic and} inelastic differential
cross sections. 
%
%
We have seen that, as a result of the strong $\beta$-$\gamma$ coupling in
transitional nuclei, the transition density for the $2_2^+$ state
exhibits structure different from those for the spherical vibrator and the prolate rotor,
which leads to the shift of the diffraction pattern of the inelastic differential
cross section for the $2_2^+$ state.
Thus, it can be a experimental signature of the strong $\beta$-$\gamma$ coupling and
large-amplitude quadrupole shape mixing in spherical-to-prolate transitional nuclei.
}
%
%

In this Letter, 
we have used a simple model to calculate the transition densities.
\KK{The model we have used is a modification of the model in
Ref. \cite{Sato2010}, and we omitted the $\beta^6\cos^2 3\gamma$ term here. 
By adding this term to the collective potential,
our model can accommodate the oblate-prolate shape coexistence,
triaxial rotor, and $\gamma$-soft limits,
which enables us to perform a similar analysis on the large-amplitude triaxial deformation dynamics.}
Moreover, it would be interesting to use the transition densities calculated microscopically 
\KS{and check the validity of our simple model.} 
One of the authors \TT{(KS)} \KK{and his collaborators} have developed a method for
microscopically determining
the 5D quadrupole collective Hamiltonian, 
the constrained Hartree--Fock--Bogoliubov plus local quasiparticle random
phase approximation 
(CHFB+LQRPA) method. 
One of the advantages of this method
is that one can take into account the contribution from the time-odd
mean field to the inertial mass unlike the widely-used cranking formula,  
and it  was successfully applied to a variety of the large-amplitude quadrupole
collective dynamics \cite{Hinohara2010b, Sato2011, Watanabe2011,
Hinohara2011a,Yoshida2011, Hinohara2011b, Hinohara2012, Sato2012}.
Microscopic calculation of the transition density with the CHFB+LQRPA
method will be reported in a future publication.
In addition, the CHFB+LQRPA method is an approximate version of the
adiabatic self-consistent collective-coordinate (ASCC) theory with
two-dimensional collective coordinate \cite{Matsuo2000, Hinohara2007}.
The ASCC theory is an advanced version of the adiabatic time-dependent 
Hartree--Fock--Bogoliubov theory,
and has been successfully used to nuclear 
structure and 
reaction studies \cite{Hinohara2008, Hinohara2009, Wen2016, Wen2017}. 
In recent studies \cite{Sato2015, Sato2017a, Sato2017b, Sato2018}, 
theoretical aspect of the ASCC theory has been highly elucidated, 
and an extension of the theory including the higher-order
contribution of the adiabatic expansion to the collective mass
has been also proposed.
The description of the transition density with the ASCC theory would be
interesting, but it remains as a future work.

\ack

The authors thank Y. Chiba for fruitful discussion.
This work was supported by JSPS KAKENHI Grant Number JP15K05087.


\end{document}